\documentclass[conference]{IEEEtran}
\usepackage{amsmath,graphicx}
\usepackage{lineno,hyperref}
\usepackage{amssymb}
\usepackage{lineno,hyperref}
\usepackage[english]{babel}
\usepackage{multirow}
\usepackage{booktabs}
\usepackage{mathtools}
\usepackage{subcaption}
\usepackage{relsize}
\usepackage{rotating}
\usepackage{makecell}

\usepackage[usenames, dvipsnames]{color}
\title{ECGNET: Learning Where to Attend for Detection of Atrial Fibrillation with Deep Visual Attention}

\author{\IEEEauthorblockN{Sajad Mousavi, Fatemeh Afghah, Abolfazl Razi}
\IEEEauthorblockA{\textit{School of Informatics, Computing and Cyber Systems}, 
{Northern Arizona  University}, Flagstaff, USA \\ \{SajadMousavi,Fatemeh.Afghah,Abolfazl.Razi\}@nau.edu}\\
\IEEEauthorblockN{U. Rajendra Acharya}
\IEEEauthorblockA{\textit{Department of Electronics and Computer Engineering, Ngee Ann Polytechnic, Singapore} \\
\textit{Department of Biomedical Engineering, School of Science and Technology, Singapore University of Social Sciences, Singapore}\\
\textit{Department of Biomedical Engineering, Faculty of Engineering, University of Malaya, Malaysia} \\
aru@np.edu.sg} 

}

\begin{document}
%\ninept
%
\maketitle
\begin{abstract} 
% \colorbox{yellow}{colorcode}\\
% \textcolor{red}{red:remove}\\
% \textcolor{blue}{blue:comments}
% \textcolor{magenta}{magenta:add}\\
% \textcolor{red}{All of arrhythmias presented in ECG signals have their underlying patterns that differentiate them from each other. Having these important information along with the whole signal is led to a better performance.}
%  \textcolor{blue}{start with a description of why AFib is one of the most complex heart conditions to detect and a brief justification for DL's superior performance in such scenarios}

The complexity of the patterns associated with Atrial Fibrillation (AF) and the high level of noise affecting these patterns have significantly limited the current signal processing and shallow machine learning approaches to get accurate AF detection results. Deep neural networks have shown to be very powerful to learn the non-linear patterns in the data. While a deep learning approach attempts to learn complex pattern related to the presence of AF in the ECG, they can benefit from knowing which parts of the signal is more important to focus during learning. 
In this paper, we introduce a two-channel deep neural network to more accurately detect AF presented in the ECG signal. The first channel takes in a preprocessed ECG signal and automatically learns where to attend for detection of AF. The second channel simultaneously takes in the preprocessed ECG signal to consider all features of entire signals. The model shows via visualization that what parts of the given ECG signal are important to attend while trying to detect atrial fibrillation. In addition, this combination significantly improves the performance of the atrial fibrillation detection (achieved a sensitivity of 99.53\%, specificity of 99.26\% and accuracy of 99.40\% on the MIT-BIH atrial fibrillation database with 5-s ECG segments.) \footnote{This material is based upon work supported by the National Science Foundation under Grant Number 1657260. Research reported in this publication was supported by the National Institute On Minority Health And Health Disparities of the National Institutes of Health under Award Number U54MD012388.}.
% \textcolor{magenta}{the abstract is not impressive, you should talk about the novelty of using attention in signal processing and how it can improve the DL performance}
% achieves a sensitivity 1.0, specificity of 1.0 and accuracy of 1.0 on some PhysioBank datasets.
\end{abstract}
\section{Introduction}
Atrial fibrillation (AF) is the most prevalent type of arrhythmia leading to hospital admissions \cite{elmoaqet2017new}, and is currently affecting the lives of more than 3 million people in the U.S. and over 33 million worldwide, while the number of AF patients in the US is expected to double by 2050 \cite{AF_2020}. Its incidence is associated with an increase in the risk of stroke, congestive heart failure, and overall mortality \cite{hamid4,hamid3}. Interpretation of ECG signals by the cardiologists and medical practitioners is usually a time-consuming task and prone to errors. Moreover, the complexity of the patterns associated with AF and the high level of noise affecting these collected signals have significantly limited the accuracy and reliability of the monitoring systems designed for AF detection.
% before it reaches AF-induced complications.
Also, the majority of these methods while have shown an effective performance on one set of data,  they may fail on other tests, discouraging the community to use any of these methods in clinical settings \cite{Ross,zaeri2018feature,afghah2015game}. 

Therefore, it is desirable to develop algorithms for automatic detection
of AF with high diagnostic accuracy and reliability. Several algorithms have been introduced to automatically detect the presence of AF based on ECG signal characteristics. Most of them depend on the detection of P-waves and R-peaks. As a result, the performance significantly degrades, if an algorithm misses detecting the relevant peaks or waves because of the presence of noise in the ECG. Although there are some research \cite{asgari2015automatic,de2006patient} that eliminate the detection of the P wave and R peak in their methodologies, they still need to extract hand-crafted features that might not be totally representative features if the dataset changes in terms of size and the presence of other arrhythmias.   

Deep learning (DL) can model high-level abstractions in data using deep networks of supervised and/or unsupervised learning algorithms, in order to learn from multiple levels of abstractions. It learns a very complex function that represents a map between inputs and targets. Over past years, deep learning based methods have been used in ECG analysis and classification. However, their performances have not been quite significant as expected them to be. Thus, developing new deep learning architectures that match specific medical problems and can capture the specific characteristics of the ECG signal is still a challenge \cite{mousavi2018inter}. 

Motivated by the aforementioned limitations, we propose an end-to-end deep visual network for automatic detection of AF called ECGNET. 
The model is a two-channel deep neural network to more accurately detect
AF presented in the ECG signal. The first channel takes in a preprocessed ECG signal and automatically learns where to attend for detection of AF.  The  second channel simultaneously takes in the preprocessed ECG signal to consider all entire signals. In this way, it gives more weights to the related parts of the ECG signal with higher potential relevance to AF, and at the same time considers the whole cycle (i.e., the beat) to extract other consecutive dependencies between each wave (i.e., P-,  QRS-, T-waves,  etc.). Moreover, the proposed approach visualize the parts of a given  ECG  signal that are more important to attend while trying to detect atrial fibrillation. It is also worth mentioning that, despite the majority of current AF detection techniques, our proposed method is capable of detecting AF in very short ECG recordings (e.g. duration around 5 s).

To the best of our knowledge, this is the first study that uses the whole information provided by the input and the visual attention at the same time for the purpose of AF detection. Recently, Shashikumar et al. \cite{shashikumar2018detection} reported an attention mechanism to detect AF, where they considered a deep recurrent neural network on 30-second ECG windows' inputs and took advantage of some time series covariates. The key contribution of our method is to develop an end-to-end two-channel deep network that automatically extract features from the focused parts of the signal with the capability of focusing on each part of a cycle (i.e., P-, QRS-, T-waves, etc.) instead of each windowed segment and from the abstracted features of the whole segment, just 5-s ECG segments. It is worth mentioning that our method does not rely on any hand-crafted features to the network as considered in \cite{shashikumar2018detection}. We also visualize which regions of the signal are important while there is an underlying AF arrhythmia in the signal. Therefore, the proposed method can potentially assist the physicians in AF detection and can be also utilized to recognize complex patterns in the signals related to other arrhythmias that cannot be easily seen in the signals. 

The rest of this paper is organized as follows. Section \ref{sec:dataset} introduces the data preparation approach and the used database in this study. Section \ref{sec:propsed} gives a
detailed description of the proposed method. Section
\ref{sec:experimental} describes the experimental setup and presents the visualization and results. Section \ref{sec:dis_con} discusses the obtained results and performance comparison to the state-of-the-art algorithms, followed by the conclusions.
% , and without any needs of the feature selection or hand-crafted feature extraction.

% \label{sec:intro}
% \textcolor{magenta}{make sure to address the advantages of this model compared to the recent literature, novelty of attention model and summary of contributions and results in the introduction.}

\begin{figure*}[htb]
\centering
  \includegraphics[width=0.9\linewidth,height=0.4\textheight,keepaspectratio]{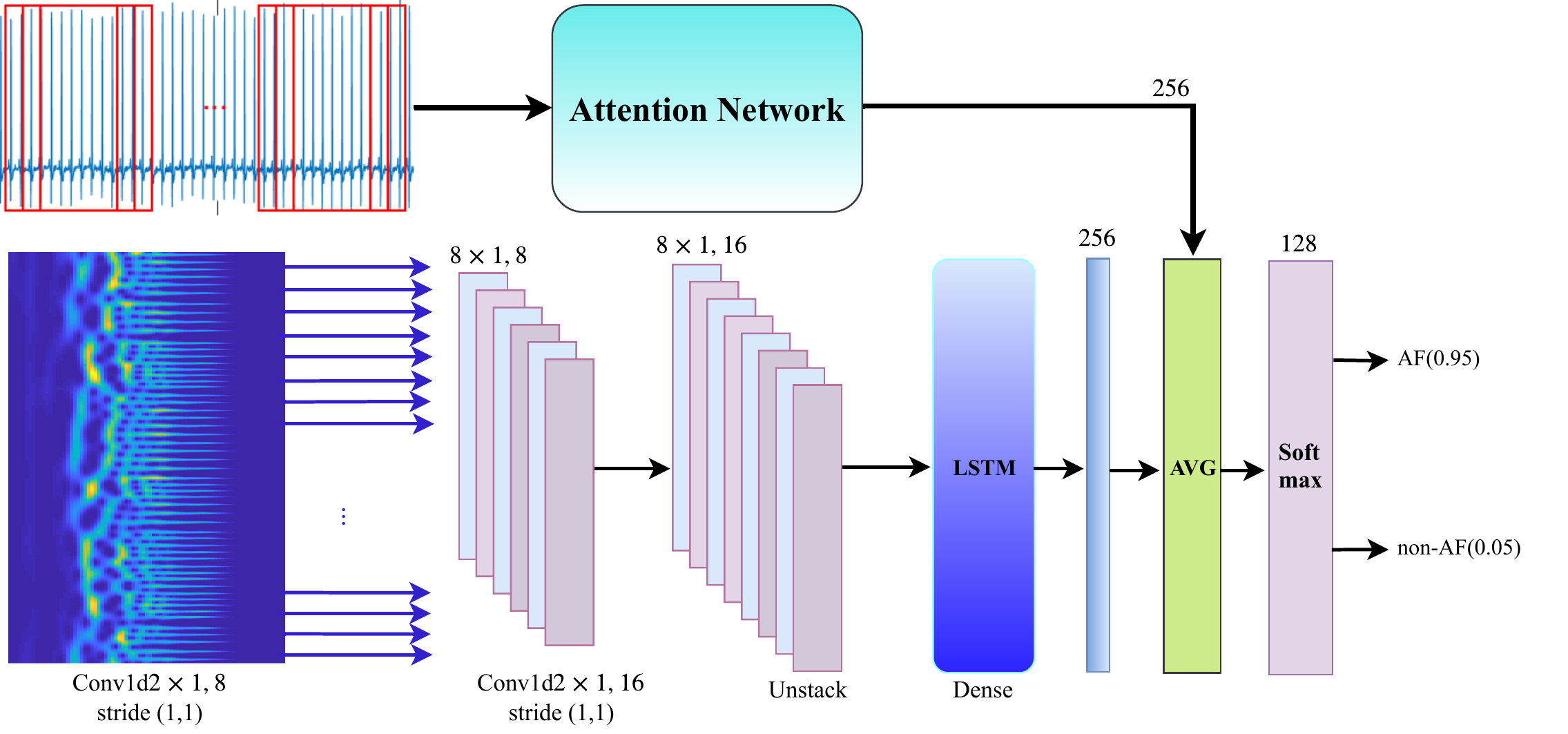}
  \caption{The network architecture for AF detection method. The top channel gets a sequence of split ECG signal (i.e., window = 128 and stride = 30) and the bottom channel gets the wavelet power spectrum of the sequence. Then, the average of two sections is computed and fed into a softmax layer. AVG: average.} 
  \label{fig:final-model}
\end{figure*}

\begin{figure}[htb]
\centering
  \includegraphics[width=\linewidth,height=0.6\textheight,keepaspectratio]{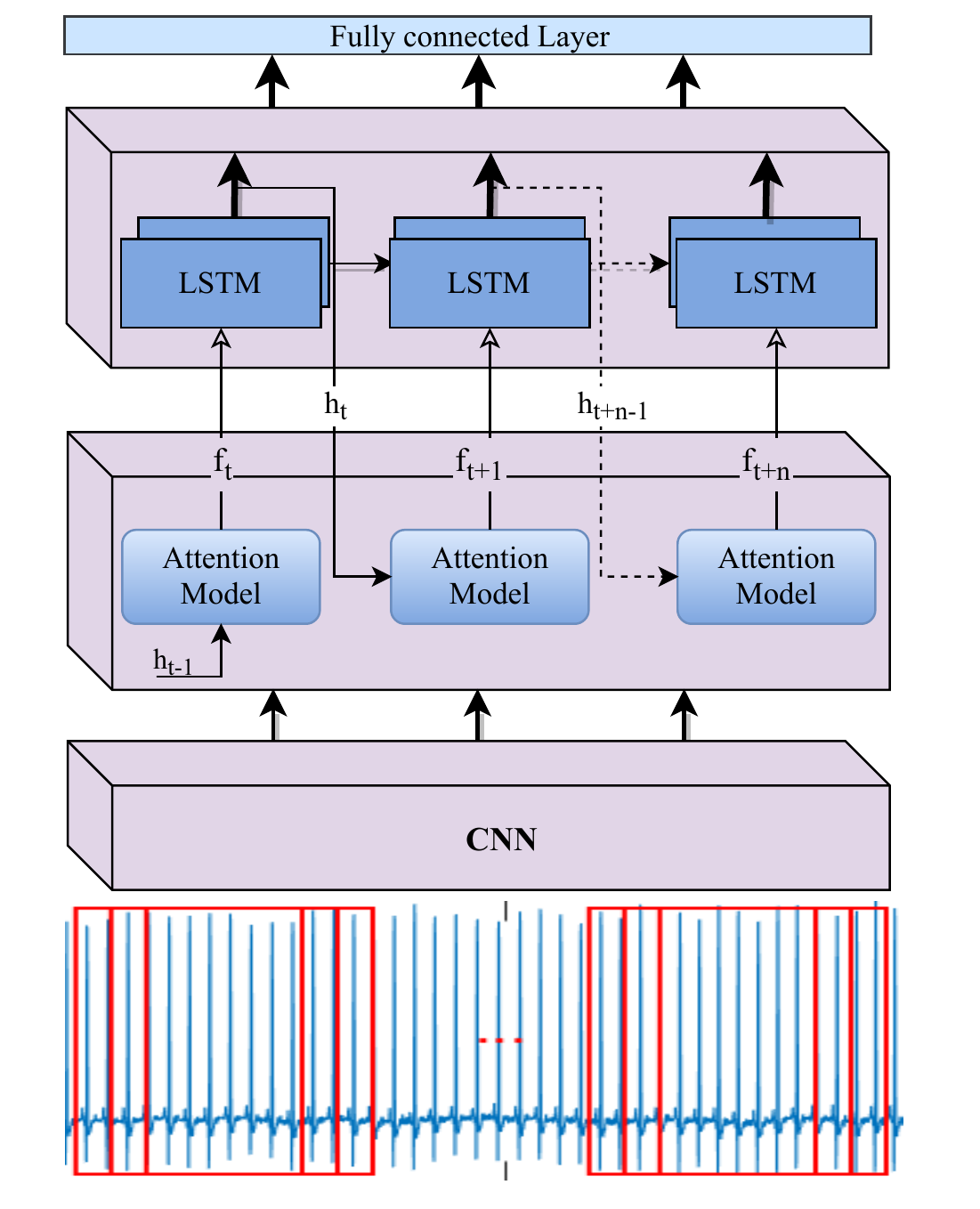}
  \caption{The network architecture of attention model.}
  \label{fig:ecg_attention}

\end{figure}

\section{Dataset and Data Preparation} 
\label{sec:dataset}
The proposed method has been evaluated using several PhysioNet databases, including  MIT-BIH Atrial Fibrillation Database (AFDB), the Long-Term Atrial Fibrillation Database (LTAFDB)
% \footnote{ http://www.physionet.org/physiobank/database/afdb/}
% \footnote{https://physionet.org/pn3/ltafdb/} 
% (NSRDB)\footnote{https://www.physionet.org/physiobank/database/nsrdb/}
and Normal Sinus Rhythm Database \cite{PhysioNet}. The AF database and the long-term AF database are comprised of 25 and 84 long-term ECG recordings of human subjects with mostly atrial fibrillation, respectively. The NSR database consists of 18 long-term ECG recordings of subjects who had no significant arrhythmias including both men and women (20 to 50 years old). The AF database includes two 10-hours long ECG recordings for each individual. The signals are sampled at 250 Hz with 12-bit resolution over a range of $\pm10$ millivolts. Individual recordings in the long-term AF database are typically 24- to 25-hours long, and sampled at 128 HZ with 12-bit resolution over a range of $20$ millivolts. 
We followed two strategies to extract the AF and non-AF excerpts of the available ECG signals. First, we extracted 30-s excerpts of the mentioned databases. To select the excerpts with AF tag, we considered rhythm annotations of type \textit{(AFIB} (atrial fibrillation) of each AF episodes. Similarly, we got a guide of rhythm annotations of type \textit{(N} (normal sinus rhythm) to select non-AF excerpts. Then, to prepare inputs for the proposed model, each ECG signal (i.e., the excerpts) of each patient was divided into a sequence of windows with lengths of 128 and 200 for the AFDB-NSRDB and LTAFDB databases, respectively, and an overlap of roughly 25\%. Second, each ECG signal of AFBD is divided into 5-s segments and each segment is labeled based on a threshold parameter, $p$. When the percentage of annotated AF beats of the 5-s segment is greater than or equal to $p$, we considered it as AF, otherwise non-AF arrhythmia. Similar to previous reported studies in \cite{xia2018detecting,asgari2015automatic}, we selected $p=50\%$. It is worth noticing that no noise removal approaches have been applied to the ECG signals.  

%  consider a section named training to talk about training and test sets of each dataset...
\section{Proposed approach}
\label{sec:propsed}
\subsection{Model Description}
An overview of the proposed model for AF detection is depicted in Fig. \ref{fig:final-model}. The model architecture is a two-channel deep neural network. The top channel takes the row windowed signal as input and includes an attention strategy to emphasize on important visual task-relevant features of the given signal. This section of the architecture is called Attention Network. We divided the given ECG signal into several windows with size 128 and an overlap of 25\%. The bottom channel considers a deep recurrent convolutional network that takes wavelet power spectrum of the windowed ECG signal. The output of the network is a vector of decimal probabilities regarding the classes. A more detailed explanation of each section of the network is provided below.

\subsection*{Attention Network}
\begin{figure}[htb]
\centering
   \includegraphics[width=6cm,keepaspectratio]{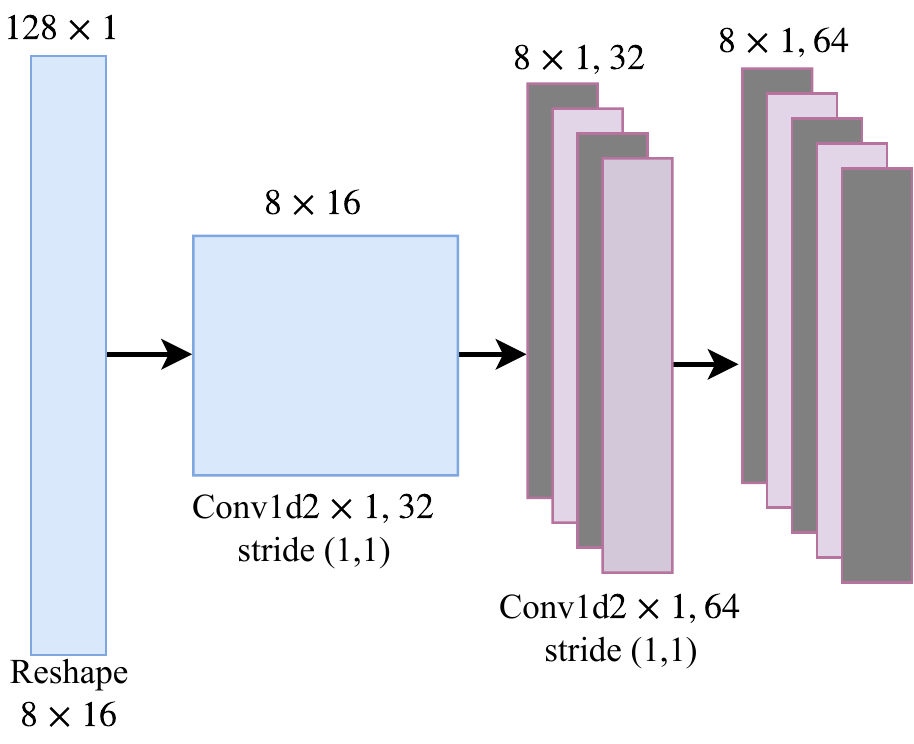}
  \caption{A diagram of convolutional layers used in the proposed model. The CNN part of the model takes the windowed ECG signal as input (i.e., a sequence of frames) and computes vertical feature slices, $F_t$ with dimension $D$.}
  \label{fig:cnn_part}

\end{figure}

% \textcolor{magenta}{so far it has been no justification for using attention and its potential advantages, I suggest to start with an intro to attention, review of its literature (if not in introduction section), differences between soft and hard attention and basic formulation before moving to explaining the architecture of your model. }
Overall, there are two types of attention models: the soft attention and the hard attention models. The soft attention models are end-to-end approaches and differentiable deterministic mechanisms
that can be learned by gradient-based methods. However, hard attention models are stochastic processes and not differentiable. Thus, they can be trained by using the REINFORCE algorithm \cite{williams1992simple} in the reinforcement learning framework \cite{mousavi2016deep}.
In this paper, a soft attention mechanism is used because the backpropagation seems to be more effective  \cite{xu2015show,mousavi2016learning}. Figure \ref{fig:ecg_attention} depicts a schematic diagram  of the Attention network. It includes three main parts as follow:

\noindent\textbf{ Convolutional neural network (CNN)}: The CNN consists of two consecutive one-dimensional convolutional layers followed by Rectified Linear Unit (ReLU) non-linearity layers. They have 32 and 64 filters of $2\times1$ with strides 1 for each one. Figure \ref{fig:cnn_part} depicts the detailed architecture.  Sequences of windowed ECG signals are fed into the CNN for feature extraction. At each time-step $t$, a windowed frame is fed into the network and the last convolutional layer of the 1-Dimensional CNN part outputs $D$ feature maps of size $K\times1$ (e.g, we concluded $64$ feature maps $8\times1$).
Then, the feature maps are converted to $K$ vectors in which each vector has $D$ dimension as follows: $$F_t = [F_{t,1},F_{t,2}, \ldots ,F_{t,K}], \qquad F_{t,i}\in \mathbb{R}^D.$$

\begin{figure}[htb]
\centering
  \includegraphics[width=7.5cm,height=0.6\textheight,keepaspectratio]{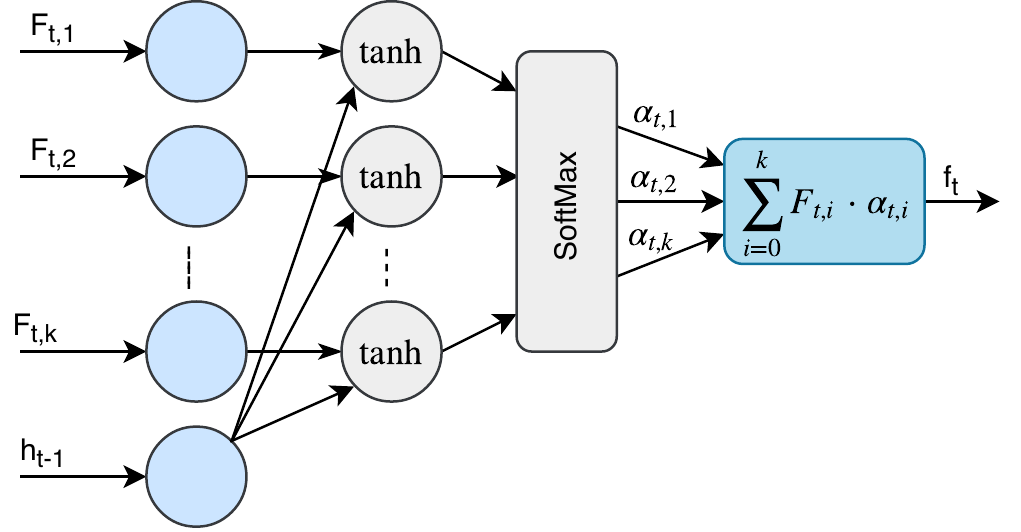}
  \caption{The structure of the attention mechanism used in the proposed model. At each time step $t$, the attention module utilizes $F_t$ and the previous hidden state of the RNN part, $h_{t-1}$ to calculate an expected value, $f_t$  with respect to  vertical feature slices, $F_{t,i}$ and the importance of each region of input window frame, $\alpha_{t,i}$.}
  \label{fig:attention_layer}

\end{figure}

\noindent\textbf{Attention layer (i.e., a soft attention mechanism)}: The extracted features of the CNN part are sequentially passed to the attention layer to compute the probabilities corresponding to the importance of each part of the windowed frame (e.g., P-, QRS- and T-waves, etc.). In other words, the input window is divided into $K$ regions and the attention mechanism attempts to attend to the most relevant regions which are related to AF. Figure \ref{fig:attention_layer} shows the structure of the attention mechanism. The attention layer gets two separate inputs: 1) $K$ vectors, $F_{t,1},F_{t,2}, \ldots ,F_{t,K}$, where each $F_{t,i}$ is a representation of different regions of the input window frame, and 2) A hidden state $h_{t-1}$, which is the internal state of the LSTM at the previous time step. Then, it computes a vector, $f_t$ which is a linear weighted combination of the values of $F_{t,i}$. Therefore, the attention mechanism can be formulated as follows:
\begin{align}
  &\begin{aligned}
    c_{att}(F_{t,i},h_{t-1}) & =\tanh({W_{h}} h_{t-1}+{W_{f}} F_{t,i}),
  \end{aligned}\\
  &\begin{aligned}
    \alpha_{t,i} & = softmax(c_{att}) \approx  \frac{\exp({c_{att}(F_{t,i},h_{t-1}}))}{\sum_{j=1}^{k} \exp({c_{att}(F_{t,j},h_{t-1}}))} \\
      & i \in 1,2,\ldots ,{k},
  \end{aligned} \\
    &\begin{aligned}
   f_t & =\sum_{i=1}^{k} \alpha_{t,i} F_{t,i},
  \end{aligned}
\end{align}
% \colorbox{yellow}{define all these parameters}
where $\alpha_{t,i}$ is the importance of the region $i$ of the input window frame.
At each time step $t$, the attention module calculates $c_{att}$, a composition of the values of $F_{t,i}$ and $h_{t-1}$ followed by a $\tanh$ layer. Then, it is passed to a softmax layer to compute $\alpha_{t,i}$ over ${k}$ regions. Indeed, each $\alpha_{t,i}$ is considered as the amount of importance of the corresponding vector $F_{t,i}$ among ${K}$ vectors in the input window. Finally, the attention layer computes $f_t$, a weighted sum of all the vectors $F_{t,i}$ based on calculated $\alpha_{t,i}$'s. Thus, the network can learn to put more emphasis on the interesting parts (e.g., P-, QRS- and T-waves, etc.) of the input window frame with higher probabilities of the presence of AF in the input ECG.
%Note that the soft attention model is fully differentiable which allows training the systems in an end-to-end manner.
  
%   the extracted features of CNN parts are sequentially passed to the attention layer to put more emphasis on the most relevant parts of input signal which are related to AF. 
\noindent\textbf{Recurrent neural network (i.e., Long Short-Term Memory (LSTM) units)}:  The attention layer is followed by LSTM units (which are a stack of two LSTM layers with the LSTM sizes of 64) for long-term learning to capture temporal dependencies between windows of each input signal. The RNN part of the network utilizes the previous hidden state $h_{t-1}$ and the output of the attention module $f_t$, to calculate the next hidden state $h_{t}$.  The parameter $h_{t}$ is used as the input of the attention module in order to calculate the value $f_{t+1}$ at the next time-step. In addition, it is utilized as the input of a fully-connected linear layer with 256 neurons. 
% followed  by a softmax classification layer with two outputs, indicating AF and non-AF classes. 

\subsection*{Deep Recurrent Convolutioal Neural Network (RCNN)}
% The first layer consists of $16$ convolution filters of size $7 \times 7$ with stride 1 followed by a Rectifed Linear Unit (ReLU) nonlinearity, a max pooling layer of size $3$ with stride 2 and a batch normalization layer. The second layer is comprised of $32$ convolution filters of size $5\times5$ with stride 1, followed by an ReLU, a max pooling layer of size $3$ with stride 2, and a batch normalization layer. The third layer is made up of $64$ convolution filters of size $3\times3$ with stride 1, again followed by a rectifier nonlinearity, a max pooling layer of size $3$ with stride 2 and a batch normalization layer. The final layer is a RNN layer with the LSTM units of size $256$.

The first layer consists of $8$ 1-D convolution filters of size $2 \times 1$ with a stride 1 followed by a Rectified Linear Unit (ReLU) non-linearity. The second layer is comprised of 16 1-D convolution filters of size $2\times1$ with stride 1, again followed by a rectifier non-linearity. The third layer is an RNN layer with the LSTM units of size $256$ followed by a fully connected layer with 256 hidden units. Here, the spectrogram size is $90\times 300 \times3$. It can be considered as a sequence of column vectors (300 vectors) that each consists of 270 values. For the purpose of feature extraction, we feed these sequences to the first 1-D convolutional layers of the deep RCNN. 

Similar to other deep learning-based AF detectors \cite{xia2018detecting,he2018automatic}, the deep neural network part of our model takes a 2-D representation with the wavelet power spectrum of the ECG segment. They employ 2-D convolution operators on the entire input, while our method applies 1-D convolution operators to each frequency vector (i.e., at each time step) of the given the spectrograms obtained from each segment, and feeds the output of the 1-D convolutional layers to long short-term memory units to capture dependencies between each frequency vector. Indeed, we consider the temporal potential patterns that might be present in the AF arrhythmia. In other words, a CNN with two-dimensional filters shares weights of the $x$ and $y$ dimensions and considers the extracted features have the same meaning apart from their locations. However, in spectrograms, the two dimensions show the strength of frequency and the time, and are completely diffident. In a 2-D convolution operator, frequency shifts of a signal (in a spectrogram representation) can change its spatial extent. Hence, the ability of 2-D CNNs to learn the spatial invariant features might not be well for the spectrograms \cite{wyse2017audio}. This is the main reason, we included 1-D CNNs followed by LSTM units instead of 2-D CNNs. Moreover, using 1-D CNNs in the network would bring a lower number of parameters and as a result further complexity reduction.

% \colorbox{yellow}{it's not clear to me and probably to the reviewers, the reasons behind selecting this architecture, you need to justify this more deeply}

Finally, the outputs of the attention and RCNN sections are averaged and fed into a softmax layer. Then, the softmax assigns decimal probabilities to each class of interest (i.e., AF and non-AF).  

% \subsection{Network Training}
% There are two ways to train the network. First,...
\begin{figure*}[t!]
\centering
  \includegraphics[keepaspectratio, width=0.7\textwidth]{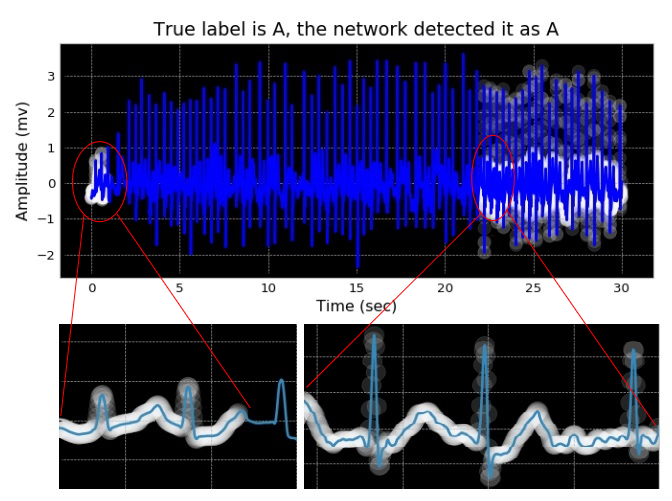}
  \caption{Visualization of the attention network's result on an ECG sample with AF arrhythmia. The white circles depicts the most important regions of the ECG signal to attend. More brightness means more attention.  }
  \label{fig:attend}

\end{figure*}  
\section{Experimental evaluation}
\label{sec:experimental}
\subsection{Experimental setup}
\label{sec:setup}
We evaluated the performance of our proposed method using different databases with various segmenting strategies.
In the first scenario, we evaluated the proposed method with 30-s ECG segments which came from two sets of data: 1) the AF database was used to extract AF samples and the NSR database was utilized to extract non-AF samples. We considered around $20,000$ samples of 30-s extracted AF and non-AF segments (each class $10,000$). $70\%$ of them were used to train the network, $10\%$ were used to validate the model, and the remaining $20\%$ were used to test the model, and 2) the Long-Term AF database was used to generate 30-s AF and non-AF excerpts. Again $20,000$ 30-s segments were considered; $70\%$ of segments were examined as the training set, $10\%$ of data were utilized for the validation set and the rest $20\%$ as the test set. For the second scenario, the AF database was utilized. There are a total of $162,536$ 5-s segments, where the number of AF segments is $61,924$, and the number of non-AF segments is $100,61$2. We randomly selected the same number of segments for each class (i.e., AF and non-AF), as $20,000$; totally $40,000$ samples, to remove the effect of imbalanced data samples on training the model. Similarly, $70\%$ of data samples were allocated to train the model, $10\%$ of them to validate the network, and $20\%$ to test the model.

The network was trained for $4,000$ steps (50 epochs) and the initial LSTM hidden and cell states were set to zero. All network weights were updated by the Momentum optimizer with mini batches of size $32$. An exponential decay function with an initial learning rate of $\alpha = 0.01$ was applied to lower the learning rate as the training progresses. Also, an additional $L_2$ regularization element with $\beta = 0.001$ was added to the loss function to reduce overfitting. The code was written in Python and Google Tensorflow which is an open-source software library for implementing the deep-learning based algorithms. To  evaluate  the  power  of  the  proposed  model  for  AF  detection  against  the  other
algorithms, four measures were considered (commonly used in the literature), which are defined as:

\begin{flalign}
&Accuracy = \frac{TP + TN}{TN+FP+FP+FN}&\\
&Sensitivity =TP/(TP+FN)& \\
&Specificity = TN/(TN+FP)&\\
&F_1\,measure = (F_{1N}+F_{1A})/2,
\end{flalign} 

where TP (True Positive), TN (True Negative), FP (False Positive) and FN (False Negative) indicate the number of excerpts correctly labeled as AF, number of excerpts correctly identified as non-AF, number of excerpts that incorrectly labeled as AF, and number of excerpts which were not identified as AF but they should have been, respectively. Also, $F_1$ measure is an average of the two $F_1$ values from each classification type (i.e., non-AF and AF) so that they can be defined as:
\begin{align*}
F_{1N} &= \frac{2\times(TN)}{(TN+FP)+(TN+FN)}, \\
F_{1A} &= \frac{2\times(TP)}{(FN+TP)+(FP+TP)}
\end{align*}

\begin{table*} [htb] 
\caption{Comparison of performance of the proposed model against other algorithms on the MIT-BIH AFIB database with the ECG segment of size 5-s ($\leq$ 7 Beats).}
 \centering{
\label{tab:compare}
	\resizebox{0.7\linewidth}{!}{  %fit to windows command 
\begin{tabular}{cccccc}
\toprule
\textbf{} & \textbf{} & \multicolumn{3}{c}{\textbf{Best Performance (\%)}} \\
\cmidrule(lr){3-5}
\textbf{Method} &  \textbf{Database} &  {$Sensitivity$} & {$Specificity$} & {$Accuracy$} \\
\midrule
ECGNET &AFDB  &$\textbf{\underline{99.53}}$ &$\textbf{\underline{99.26}}$ &$\textbf{\underline{99.40}}$  \\
% Cui, et al. (2017) \cite{cui2017automated} &AFDB  &$97.04$ &$97.04$ &$97.78$  \\
Xia, et al. (2018) \cite{xia2018detecting} &AFDB  &$98.79$ &$97.87$ &$98.63$  \\
Asgari, et al. (2015) \cite{asgari2015automatic} &AFDB  &$97.00$ &$97.10$ &$-$  \\
Lee, et al. (2013) \cite{lee2013atrial} &AFDB  &$98.20$ &$97.70$ &$-$  \\
Jiang, et al. (2012) \cite{jiang2012high} &AFDB  &$98.20$ &$97.50$ &$-$  \\
Huang, et al. (2011) \cite{huang2011novel} &AFDB  &$96.10$ &$98.10$ &$-$  \\
Babaeizadeh, et al. (2009) \cite{babaeizadeh2009improvements} &AFDB  &$92.00$ &$95.50$ &$-$  \\
Dash, et al. (2009) \cite{dash2009automatic} &AFDB  &$94.40$ &$95.10$ &$-$  \\
Tateno, et al. (2001) \cite{tateno2001automatic} &AFDB  &$94.40$ &$97.20$ &$-$  \\
 \bottomrule  
\end{tabular} }
}
\end{table*}

\begin{table} [ht]  

\caption{Confusion matrix achieved by the proposed method on the MIT-BIH AFIB database with ECG segment of size 5-s (
$\leq$ 7 Beats).}
\renewcommand{\arraystretch}{1.2}
 \centering{
\label{tab:cm-5s}
	\resizebox{0.8\linewidth}{!}{  %fit to windows command 
\begin{tabular}{cccccc}
 \toprule
 &  &  \multicolumn{2}{c} {Predicted}& Total \\
 \cmidrule(lr){3-4} 
\cmidrule(lr){5-5}
 & & non-AF & AF \\
% \midrule
\cline{2-5}
\multirow{3}{*}{\begin{turn}{90}Actual\end{turn}}  &\multicolumn{1}{|l|}{ non-AF}
& 3891 &  29    &  3920  
 \\ 
&\multicolumn{1}{|l|}{AF}
&  19 & 4061   &  4080   
 \\ 
 \cline{2-5}
&\multicolumn{1}{|l|}{Total}
&  3910 & 4090   &  8000   
 \\ 
  \bottomrule  
\end{tabular} }
}
\end{table}

\subsection{Visualization and Results}
\label{sec:visual_results}
 Table \ref{tab:compare} presents a performance comparison between the proposed method and several state-of-the-art algorithms using the MIT-BIH AFIB database, where the 5-s data segments were considered. As it is clear in Table \ref{tab:compare}, overall, our proposed AF detector shows better results in terms of the sensitivity, specificity and accuracy evaluation metrics compared to all methods presented in the table.
 
 Tables \ref{tab:cm-5s} presents the confusion matrix achieved by the proposed method on the MIT-BIH AFIB database with ECG segment of size 5-s ($\leq$ 7 Beats). The first cell of the diagonal of the confusion matrix denotes the number of excerpts correctly identified as non-AF (the true negative), the second cell indicates the number of excerpts correctly labeled as AF (the true positive). As shown in the table, the true/negative values are much greater than the other values in their corresponding rows.

 To further evaluate the performance of this method, we also calculated the $F_1$ score of the detection outcome for different databases. Table \ref{tab:f1scores} summarizes the performance of proposed model with different data segmentation approach. We should note that the $F_1$ scores of other algorithms in the literature were not available to compare with. In addition, to consider the 5-s data segments as input to the network, we employed data segments of 30 seconds in our experiments. Table \ref{tab:compare_3models} reports the results of sensitivity, specificity, and accuracy of the proposed detection models. As it is shown in the table, our models result in the best possible performances with sensitivity, specificity and accuracy of $1.0$ on the MIT-BIH AFIB dataset as well as stunning outcomes on the Long-Term AFIB dataset.     

A visualization example of attended parts of an ECG signal with an AF is illustrated in Figure \ref{fig:attend}. The white regions, showing with circles indicate where the model learned to look while the patient had the  AF. We should note that the two main indicators of AF in ECG signals as considered in the majority of the previous works are: 1) the absence of P-waves that can be replaced by a series of low-amplitude oscillations called fibrillatory waves, and 2) the irregularly irregular rhythm (i.e., irregularity of R-R intervals) \cite{jiang2012high,babaeizadeh2009improvements,huang2011novel}. It is worth noting that the attention network focused on the regions where there is no P-waves or fibrillatory waves in the ECG signal. Furthermore, there are some attentions on the R-peaks that may show the irregularity of R-R intervals. As the R-R intervals of the signal were not computed here, we do not have a reference point to show if the focused R-peaks are because of the irregularity. We encourage interested readers to view the clip video through the link below that visualizes the attended locations by the proposed model as well.
% In the MIT-BIH database, although records 04936 and 05091 contain incorrect annotations, we did not exclude them....
Video: \url{https://drive.google.com/open?id=1VMhoVPzUVsXjP3JSge9_PJGAUImZTinN}. As shown in Tables \ref{tab:compare},\ref{tab:f1scores} and \ref{tab:compare_3models}, our attention-based DL model with the proposed combination architecture of the attention network and the deep recurrent neural network (i.e., deep net part of the entire model) has superior performance in detection of AF compared to the existing techniques. 
% the a the gaining maximum benefit from  combined the same time 

\begin{table*} [htb] 
\caption{$F_1$ scores of each classification type obtained by the proposed method on different PhysioNet databases.}
 \centering{
\label{tab:f1scores}
	\resizebox{0.9\linewidth}{!}{  %fit to windows command 
\begin{tabular}{ccccccc}
\toprule
\textbf{} & \textbf{} & \textbf{} & \multicolumn{3}{c}{\textbf{Performance (\%)}} \\
\cmidrule(lr){4-6}
\textbf{Method} &  \textbf{Database}&\textbf{Length} &  {$F_{1N}$} & {$F_{1A}$} & {$F_1$} \\
\midrule
Proposed Method &AFDB  &5-s ($\leq$ 7 Beats)&$99.40$ &$99.39$ &$99.40$  \\
Proposed Method &AFDB-NSRDB  &30-s ($\approx$ 45 Beats)&$1.0$ &$1.0$ &$1.0$  \\
Proposed Method &LTAFDB  &30-s ($\approx$ 37 Beats)&$98.40$ &$98.50$ &$98.45$  \\
 \bottomrule  

\end{tabular} }
}
\end{table*}

\begin{table*}  
\caption{Detection performances of the proposed method on different PhysioNet databases with the ECG segment of size 30-s.}
 \centering{
\label{tab:compare_3models}
	\resizebox{0.9\linewidth}{!}{  %fit to windows command 
\begin{tabular}{ccccccc}
\toprule
\textbf{} & \textbf{} &\textbf{} & \multicolumn{3}{c}{\textbf{Best Performance (\%)}} \\
\cmidrule(lr){4-6}
\textbf{Method} & \textbf{Database} & \textbf{Length} &  {\textbf{Sensitivity}} & {\textbf{Specificity}} & {\textbf{Accuracy}} \\
\midrule
The entire network &AFDB-NSRDB & 30-s ($\approx$ 45 Beats) &$1.0$ &$1.0$ &$1.0$  \\
Attention Network &AFDB-NSRDB & 30-s ($\approx$ 45 Beats) &$1.0$ &$1.0$ &$1.0$  \\
Deep CNN part &AFDB-NSRDB & 30-s ($\approx$ 45 Beats) &$1.0$ &$1.0$ &$1.0$  \\

The entire network &LTAFDB & 30-s ($\approx$ 37 Beats) &$99.85$ &$99.90$ &$99.91$  \\
Attention Network &LTAFDB & 30-s ($\approx$ 37 Beats) &$1.0$ &$99.80$ &$99.89$  \\
Deep RCNN part &LTAFDB & 30-s ($\approx$ 37 Beats) &$99.35$ & $99.45$ &$99.41$  \\
\bottomrule  
\end{tabular} }
}
\end{table*}

\section{Discussion and Conclusion}
\label{sec:dis_con}
Several algorithms have been reported in the literature to detect AF  ranging from RR interval variability, P-wave detection based to machine learning paradigms including deep learning-based methods. In this paper, we propose an attention-based AF detection method that automatically identifies the potential AF regions in the ECG signal.
Despite the majority of the previously reported works in this area \cite{huang2011novel,lee2013atrial,babaeizadeh2009improvements,jiang2012high}, the performance of the proposed method does not rely on the accuracy of the hand-crafted algorithms for detecting P-waves and R-R intervals. In contrast, the attention network in our proposed model automatically focuses on the most relevant regions of each heartbeat which is prone to be a part of an AF and puts more weights on those regions in the network to distinguish between the AF and non-AF classes. The performance of our method is compared against the majority of existing algorithms for detection of  AF using the same databases and the same evaluation metrics. The proposed method achieved an accuracy of $99.40$, a sensitivity of $99.53$ and a specificity of $99.26$ validated on the MIT-BIH AFIB database with 5-s data segments that significantly outperforms the results of other studies. Moreover, the detector obtained the value $1.0$ for the accuracy, sensitivity and specificity on the MIT-BIH AFIB database with 30-s ($\approx$ 45 Beats) data segments, and the values $99.91$, $99.85$ and $99.90\%$ for the accuracy, sensitivity and specificity, respectively on the Long-Term AFIB database with 30-s ($\approx$ 37 Beats) data segments.

One of the other key issues in AF detection methods is their poor performance in detecting AF episodes in short signal recordings (i.e., less than 30-s).
 While the majority of the state-of-the-art algorithms require a 30-s episode or at least 127/128 beats to achieve an acceptable detection performance \cite{cui2017automated,zhou2014automatic,huang2011novel,lee2013time}, our proposed method offers great performance on very short ECG segments of size 5-s which inhere are less than a 7-beats window. 

Thanks to the proposed end-to-end deep learning approach, we do not need tuning any parameters which might affect the detection performance. For instance, the setting of parameters in \cite{cui2017automated,petrenas2015low} have a significant influence on optimizing their final results. 

In this study, we proposed a novel deep network architecture to classify the given signal as AF or non-AF. The proposed AF detector shows better performance compared to the other detectors in the literature. One key aspect of our AF detector is that it simultaneously gives more weights to the related parts of the ECG signal with higher potential prevalence of AF, and also considers the whole cycle (i.e., the beat) to extract other consecutive dependencies between each wave (i.e., P-, QRS-, T-waves, etc.). Moreover, The proposed method obtains significant detection results by using a short ECG segment, 5-s long, to detect AF without the need for tuning any parameters in the model. 

% Below is an example of how to insert images. Delete the ``\vspace'' line,
% uncomment the preceding line ``\centerline...'' and replace ``imageX.ps''
% with a suitable PostScript file name.
% -------------------------------------------------------------------------
% \begin{figure}[htb]

% \begin{minipage}[b]{1.0\linewidth}
%   \centering
%   \centerline{\includegraphics[width=8.5cm]{images/ECG_attention.pdf}}
% %  \vspace{2.0cm}
%   \centerline{(a) Result 1}\medskip
% \end{minipage}
% %
% \begin{minipage}[b]{.48\linewidth}
%   \centering
%   \centerline{\includegraphics[width=4.0cm]{images/cnn_part.pdf}}
% %  \vspace{1.5cm}
%   \centerline{(b) Results 3}\medskip
% \end{minipage}
% \hfill
% \begin{minipage}[b]{0.48\linewidth}
%   \centering
%   \centerline{\includegraphics[width=4.0cm]{images/attention_layer.pdf}}
% %  \vspace{1.5cm}
%   \centerline{(c) Result 4}\medskip
% \end{minipage}
% %
% \caption{Example of placing a figure with experimental results.}
% \label{fig:res}
% %
% \end{figure}

%\vfill
%\pagebreak

% \newpage
\bibliographystyle{IEEEbib}
\bibliography{ECGNET_AF}

\end{document}